\documentclass[letterpaper,11pt]{article}
\usepackage[cmex10]{amsmath}
\usepackage{amssymb,mathrsfs,amsthm}
\usepackage{float}
\usepackage{framed}
\usepackage{enumerate}
\usepackage{color}
\usepackage{xspace}
\usepackage[normalem]{ulem}
\usepackage[margin=1in]{geometry}
\usepackage{graphicx}
\usepackage{tikz}
\definecolor{DarkBlue}{rgb}{0.1,0.1,0.5}
\usepackage[colorlinks=true,linkcolor=black,citecolor=DarkBlue,urlcolor=DarkBlue]{hyperref}
\usepackage[utf8]{inputenc}  

\newtheorem{theorem}{Theorem}[section]

\newtheorem{definition}[theorem]{Definition}

\newtheorem{proposition}[theorem]{Proposition}

\renewcommand{\H}{{\cal H}}
\newcommand{\dens}{D}
\newcommand{\Hmin}{\mathrm{H}_{\min}}
\newcommand{\Guess}{\mathrm{Guess}}
\newcommand{\ket}[1]{| #1 \rangle}

\newcommand{\ketbra}[2]{|#1\rangle \!\langle #2|}
\newcommand{\proj}[1]{\ketbra{#1}{#1}}

\newcommand{\tr}{\mathrm{tr}}

\newcommand{\C}{\mathbb{C}}
\newcommand{\R}{\mathbb{R}}
\newcommand{\N}{\mathbb{N}}

\newcommand{\A}{\mathbb{A}}
\newcommand{\B}{\mathbb{B}}
\newcommand{\X}{\mathbb{X}}
\newcommand{\Y}{\mathbb{Y}}

\newcommand{\bx}{\b{x}}
\newcommand{\by}{\b{y}}
\newcommand{\ba}{\b{a}}
\newcommand{\bb}{\b{b}}

\newcommand{\bX}{\b{X}}
\newcommand{\bY}{\b{Y}}
\newcommand{\bA}{\b{A}}
\newcommand{\bB}{\b{B}}

\newcommand{\D}{\b{\frak{D}}}

\newcommand{\set}[1]{\{#1\}}

\newcommand{\bat}{\set{0,1}} 
\newcommand{\hmin}{\Hmin}

\newcommand{\ca}{{\cal A}}
\newcommand{\cb}{{\cal B}}
\newcommand{\ce}{{\cal E}}

\newcommand{\etal}{{et~al.\@}\xspace}  

\newcommand{\h}{\mathrm{h}_\circ}	
\newcommand{\clh}{\mathrm{h}}		
\newcommand{\Hist}{H\hspace{-0.1ex}ist}
\newcommand{\hist}{hist}

\newcommand{\Imax}{I_{\max}}

\renewcommand{\b}[1]{\text{\boldmath$#1$\unboldmath}}
\newcommand{\eps}{\varepsilon}

\newcommand{\BAD}{\mathcal{B}}
\newcommand{\GOOD}{\mathcal{G}}
\newcommand{\BADGUESS}{\mathcal{B}^{\mathrm{guess}}}
\newcommand{\GOODGUESS}{\mathcal{G}^{\mathrm{guess}}}

\usepackage{svn-multi}
\svnidlong{$HeadURL: $} {$LastChangedDate:$} {$LastChangedRevision: 152 $} {$LastChangedBy: $}

\begin{document}

\title{Security and Composability of Randomness Expansion \\ from Bell
  Inequalities
}

\date{\today 
}

\author{
Serge Fehr\thanks{Centrum Wiskunde \& Informatica (CWI), Amsterdam, The Netherlands.
\texttt{Serge.Fehr@cwi.nl}.} 
\and 
Ran Gelles\thanks{Department of Computer Science, UCLA, Los Angeles, CA, USA.  \texttt{gelles@cs.ucla.edu}.}
\and 
Christian Schaffner\thanks{University of Amsterdam and Centrum
  Wiskunde \& Informatica (CWI), Amsterdam, The Netherlands. 
  \texttt{c.schaffner@uva.nl}.}
}
\maketitle

\begin{abstract}
The nonlocal behavior of quantum mechanics can be used to generate guaranteed fresh randomness from an {\em untrusted} device that consists of two nonsignalling components; since the generation process requires some initial fresh randomness to act as a catalyst, one also speaks of randomness {\em expansion}.

Colbeck and Kent proposed the first method for generating randomness from untrusted devices, however, without providing a rigorous analysis. This was addressed subsequently by Pironio~\etal [\emph{Nature} \textbf{464} (2010)], who aimed at deriving a lower bound on the min-entropy of the data extracted from an untrusted device, based only on the observed non-local behavior of the device.
Although that article succeeded in developing important tools towards the acquired goal, it failed in putting the tools together in a rigorous and correct way, and the given formal claim on the guaranteed amount of min-entropy needs to be revisited.

In this paper we show how to combine the tools provided by Pironio~\etal, as to obtain a meaningful and correct lower bound on the min-entropy of the data produced by an untrusted device, based on the observed non-local behavior of the device. Our main result confirms the essence of the improperly formulated claims of Pironio~\etal, and puts them on solid ground.

We also address the question of composability and show that different untrusted devices can be composed in an alternating manner under the assumption that they are not entangled. This enables for superpolynomial randomness expansion based on two untrusted yet unentangled devices. 
\end{abstract}


\section{Introduction}

\paragraph{Background.} 

One of the counter-intuitive features of quantum mechanics is its {\em non-locality}: measuring possibly far apart quantum systems in randomly selected bases (chosen out of some given class) may lead to correlations that are impossible to obtain classically. 
Anticipated by Einstein, Rosen and Podolsky~\cite{EPR35}, it was John Bell~\cite{Bell64}  who put this property on firm ground by proposing an inequality that is satisfied by any classical correlation, but is violated when the correlation is obtained from measuring entangled quantum states. Such inequalities are called \emph{Bell inequalities}.

An important example of such a Bell inequality was proposed by Clauser
Horne, Shimony and Holt~\cite{CHSH69} and states that if $X$~and~$Y$ are independent uniformly distributed bits, and if the bit~$A$ is obtained by ``processing"~$X$ without knowing~$Y$, and the bit~$B$ is obtained by ``processing"~$Y$ without knowing~$X$, then the probability that $A \oplus B = X \wedge Y$ is at most $75\%$. This bound on the probability holds if the processing is done classically with shared randomness, but can be violated when the processing involves measuring an entangled quantum state; in this latter case, a probability of roughly $85\%$ can be achieved. 

Violating a Bell inequality necessarily means that there must be some amount of fresh randomness in the outputs $A$ and $B$ (given the inputs $X$ and $Y$). 
More formally, consider an untrusted device $\D$, prepared by an
adversarial manufacturer Eve. The device consists of two components,
set up by Eve, which on respective inputs $X$ and $Y$ produce 
respective outputs $A$ and $B$ {\em without communicating}. No matter
how the two components work, as long as a given Bell inequality is
violated during $n$ sequential interactions with $\D$ (which can be observed by
doing statistics), there must be a certain amount of uncertainty in the
$n$~output pairs $(A_1,B_1),\ldots,(A_n,B_n)$, even given the $n$~input pairs $(X_1,Y_1),\ldots,(X_n,Y_n)$, and thus it should be possible to apply a randomness extractor to obtain nearly-random bits. 

This kind of randomness expansion from untrusted devices was first suggested by Colbeck~\cite{Colbeck09} and Colbeck and Kent~\cite{CK11}, who presented a scheme that uses GHZ states and reaches a linear expansion, however, without providing a rigorous security analysis. The main point missing in these works is a method 
to rigorously bound the min-entropy of a device's output. 
The work of Pironio~\etal~\cite{nature10} addresses this issue, and they propose a technique to numerically compute a lower bound on the min-entropy of the output pair $AB$ (conditioned on $X$ and $Y$) as a function of the Bell value of the device $\D$ (which quantifies the violation of Bell inequality). For the special case of CHSH, they also show an analytical bound. 

The authors of~\cite{nature10} also consider the case of $n$
sequential interactions with $\D$, and they show how to {\em estimate}
the average Bell value of $\D$ over the $n$ rounds by doing statistics
over the observed data. This is non-trivial because the Bell value of
$\D$ may change over the different rounds, and, for each round, it may
depend on the behavior of the previous rounds. In other words, the
Bell value of $\D$ during round $i+1$ depends on the history
$(A_1,B_1,X_1,Y_1),\ldots,(A_{i},B_{i},X_{i},Y_{i})$.  Combining
things, Pironio~\etal then claim to have a bound on the min-entropy of
$(A_1,B_1),\ldots,(A_n,B_n)$, conditioned on
$(X_1,Y_1),\ldots,(X_n,Y_n)$, as a function of the observed data,
i.e., as a function of
$(A_1,B_1,X_1,Y_1),\ldots,(A_n,B_n,X_n,Y_n)$. However, such a
statement does not make sense, since the considered min-entropy is a
value determined by the experiment description (which specifies the
probability distribution), whereas the claimed bound depends on the
specific outcome of the experiment.%
\footnote{This is like saying that the min-entropy of throwing a fair
  die is lower bounded by the result of the throw: the former equals
  $\log(6) \approx 2.6$, whereas the latter is a random number in
  $\set{1,\ldots,6}$. Trying to bound the min-entropy {\em conditioned
    on the observed outcome} makes no sense either, because this
  conditional min-entropy obviously vanishes. }  Furthermore, not only
is the claim improperly formulated, but there is also a flaw in its
derivation, which is without an obvious fix.  Thus, even though the
necessary tools are provided in~\cite{nature10}, they are not put
together in the right rigorous way to be able to control the
min-entropy of $(A_1,B_1),\ldots,(A_n,B_n)$ produced by an untrusted
device $\D$.

\paragraph{Our Result.} 

In this paper, we make up for this shortfall in~\cite{nature10}. Specifically, we show how to rigorously and correctly put together the tools provided in~\cite{nature10} in order to obtain a meaninful (and correct) bound on the min-entropy of $(A_1,B_1),\ldots,(A_n,B_n)$, conditioned on $(X_1,Y_1),\ldots,(X_n,Y_n)$, by means of the observed data. The trick is to consider and bound the min-entropy {\em conditioned} on the event that the estimator for the average Bell value lies in some interval. This gives us some control over the average Bell value of the device, but, as we show, still leaves enough uncertainty in the data to get a good bound on its min-entropy. 

We also address the question of the composability of untrusted devices, and we show that under the assumption that different devices are not entangled, the output of one device, after privacy amplification, can be used as input for a second device, and the resulting output of the second device, after privacy amplification, can again be fed into the first device, etc. 
Using an extractor with a short seed 
for doing the privacy amplification, this allows for a superpolynomial randomness-expansion scheme using two untrusted (but guaranteed-to-be unentangled) devices.

\paragraph{Concurrent and Related Work.}
In concurrent and independent work, Vazirani and
Vidick~\cite{VV11} as well as Pironio and Massar~\cite{PM11} came up
with results that are overlapping with ours. We briefly discuss here
the similarities and the differences between our results and those of
Vazirani and Vidick and of Pironio and Massar. We encourage the reader
to also look at the comparisons given in~\cite{VV11,PM11}.

Vazirani and Vidick obtain a randomness-expansion scheme with
superpolynomial expansion and security against quantum side
information. We do not achieve security against quantum side
information, and our superpolynomial randomness-expansion scheme
requires two unentangled devices in an iterative way, whereas their
scheme works with just {\em one} single device.  On the other hand,
their result is tailored to CHSH and requires an almost full violation
of Bell inequality, while our result is generic and holds for any Bell
inequality, and we show that {\em any} violation leads to some amount
of fresh randomness.

Pironio and Massar's results on the other hand are very similar to
ours, and only differ in some minor details.\footnote{As a historical
  note, previous versions of their~\cite{PM11v1} and our
  paper~\cite{FGS11v2} claimed security against quantum side
  information, but both proofs were incorrect.}

In a very recent preprint, Barrett, Colbeck and Kent point out the
possibility of Trojan-horse attacks on device-independent
randomness-expansion protocols~\cite[Appendix]{BCK12}. It seems
impossible to prevent that Eve programs devices (that are used
multiple times) to release in later rounds information about previous
outputs. 
We note that although such an attack seems unavoidable,
in a \emph{single} activation of our randomness-expansion scheme
(see~\cite[Section~5]{FGS11v2} for details), 
we can re-use the same devices over and over again and
still prevent such a Trojan-horse attack by only
releasing the output of the very last round (and aborting if things go
wrong before the last round is reached).


\section{Preliminaries}

We assume the reader is familiar with quantum information processing, and we merely fix our notation and some basic concepts in this section. Throughout the paper, all logarithms are base~$2$.

\subsection{Quantum States}

The {\em state} of a quantum system $\ca$ is given by a {\em density
  matrix} $\rho_\ca$, i.e., a positive-semidefinite trace-1 matrix
acting on some Hilbert space $\H_\ca$. We denote the set of all such
matrices, acting on $\H_\ca$, by $ \dens(\H_\ca)$.  The state space of
the joint quantum systems $\ca \cb$, which consist of two (or more)
subsystems $\ca$ and~$\cb$ , is given by the tensor product $\H_{\ca
  \cb} = \H_\ca \otimes \H_\cb$. If the state of the joint system is
given by $\rho_{\ca \cb}$, then the state of the sub-system $\ca$ when
considered as a “stand alone” system is given by the reduced density
matrix $\rho_\ca = \tr_\cb (\rho_{\ca \cb}) \in
\dens(\H_\ca)$, obtained by tracing out system $\cb$.

A random variable $X$ over a finite set $\X$ with probability
distribution $P_X$ can be represented by means of the density matrix
as $\rho_X = \sum_x P_X(x) \proj{x} \in \dens(\H_X)$, where
$\set{\ket{x}}_{x \in \X}$ forms a basis of $\H_X = \C^{|\X|}$. Thus,
we may view $X$ as a quantum system, and we say that its state,
$\rho_X$, is {\em classical}. If the state of a quantum system $\ce$
depends on the random variable $X$, in that the state of $\ce$ is
given by $\rho_\ce^x \in \dens(\H_\ce)$ if $X = x$, then we can view
the pair $X \ce$ as a bi-partite quantum system in state $\rho_{X \ce}
= \sum_x P_X(x) \proj{x} \otimes \rho^x_\ce \in \dens(\H_X \otimes
\H_\ce)$. This naturally extends to multiple random variables and
quantum systems.

The distance between two states $\rho_\ce,\tilde{\rho}_\ce \in
\dens(\H_\ce)$ is measured by their {\em trace distance} $\frac12{\|
  \rho_\ce - \tilde{\rho}_\ce \|_1}$, where $\| \cdot \|_1$ is the
$L_1$ norm.%
\footnote{Defined by $\| A \|_1 := \tr(\sqrt{A^\dagger A})$, where
  $A^\dagger$ denotes the Hermitian transpose.}  In case of classical
states $\rho_X$ and $\tilde{\rho}_X$, corresponding to distributions
$P_X$ and $\tilde{P}_X$, the trace distance coincides with the
statistical distance $\frac12\sum_x |P_X(x)-\tilde{P}_X(x)|$.

\subsection{Closeness to Uniform, Min-Entropy, and Extractors}

In the following definitions, we consider a bi-partite system $X\ce$ with classical $X$, given by $\rho_{X \ce}$.
$X$ is said to be {\em random and independent} of $\ce$ if $\rho_{X\ce} = \rho_U \otimes \rho_{\ce}$, where $\rho_U$ is the fully mixed state on $\H_X$ (i.e., $U$ is classical and, as random variable, uniformly distributed). 

\begin{definition}
\label{def:distuni}
The {\em distance to uniform} of $X$ given $\ce$ is  
\(
d(X\mid\ce) : = \tfrac12 \| \rho_{X\ce} - \rho_U \otimes \rho_{\ce} \|_1.
\)
\end{definition}
If $\Omega$ is some event, determined by the random variable $X$, then $d(X\mid\ce,\Omega)$ is naturally defined by means of replacing the distribution $P_X$ by $P_{X|\Omega}$. The same applies to the next two definitions. 

\begin{definition}
The \emph{guessing probability} of $X$ given $\ce$ is  
\[
\Guess(X\mid \ce):= \sup_{\set{M_x}_x} \sum_x P_X(x) \mathrm{tr}(M_x\, \rho_{\ce}^x),
\]
where the supremum is over all POVMs $\set{M_x}_x$ on $\H_{\ce}$.
\end{definition}

\begin{definition}
The {\em min-entropy} of $X$ given $\ce$ is  given by 
\(
\hmin(X \mid\ce) := -\log \Guess(X\mid\ce).
\)
\end{definition}
This definition was shown in~\cite{KRS09} to coincide with the definition
originally introduced by Renner \cite{Renner05} which also coincides
with the classical definition of conditional min-entropy, in the case
where $\ce$ is classical.

\begin{definition}
A function $\mathsf{Ext}:\bat^n \times \bat^d \rightarrow \bat^\xi$ is a 
\emph{$(k,\varepsilon_{ext})$-strong extractor}, if for
any bipartite quantum system $X\ce$ with classical $X$ and with $\hmin(X\mid\ce) \geq k$, and for a uniform and independent seed $Y$, we have
\(
d\big(\mathsf{Ext}(X,Y)\  \big\vert \ Y \ce \big) \leq \varepsilon_{ext} \, .
\)
\end{definition}
Note that we find ``extractor against quantum adversaries" a too cumbersome terminology; thus we just call $\mathsf{Ext}$ a (strong) extractor, even though it is a stronger notion than the standard notion of a (strong) extractor.

\subsection{Bell-Inequality and CHSH}

For given finite sets $\A,\B,\X,\Y$, consider a conditional probability distribution $P_{AB|XY}$, specified as follows. There exists $\rho_{\ca\cb} \in \dens(\H_{\ca} \!\otimes\! \H_\cb)$ for an arbitrary (finite) dimensional two-partite quantum system $\ca \cb$, and families of measurements $\{M_x^a\}$ and $\{N_y^b\}$, indexed by $x \in \X$ and $y \in \Y$, acting on $\ca$ and $\cb$, and with measurement outcomes $a \in \A$ and $b \in \B$, respectively
such that $P_{AB\mid XY}(a,b\mid x,y) = \tr\bigl( (M_x^a \otimes
N_y^b) \, \rho_{\ca\cb} \, ({M_x^a} \otimes {N_y^b})^\dag \bigr)$ for all $(a,b,x,y) \in \A \times \B \times \X \times \Y$.

\begin{definition}[Bell Value]
For any set ${\cal C} = \set{c_{abxy}}$ of {\em Bell coefficients}, the {\em Bell value} of~$P_{AB|XY}$ (with respect to $\cal C$) is defined as
$$
I(P_{AB|XY}) = \sum_{abxy}c_{abxy}P_{AB\mid XY}(a,b\mid x,y)
$$
\end{definition}
$P_{AB|XY}$ is called {\em classical} (or {\em local}) if there exist
(conditional) probability distributions $P_R$, $P_{A|XR}$ and
$P_{B|YR}$ such that $P_{AB|XY}(a,b\mid x,y) = \sum_r P_R(r)
P_{A|XR}(a\mid x,r) P_{B|YR}(b\mid y,r)$ for all $a,b,x,y$; this is equivalent
to requiring that $P_{AB|XY}$ can be specified by means of a {\em
  separable state} $\rho_{\ca\cb}$.  We let $I_0$ denote the maximal
Bell value achievable (for a given set of Bell coefficients) with a
classical $P_{AB|XY}$.  We speak of a {\em violation of Bell
  inequality} if there exists a quantum system resulting in
conditional probability distribution with a Bell value greater than
$I_0$.

For instance, for so-called CHSH Bell coefficients~\cite{CHSH69}, given by $c_{abxy}=(-1)^{xy}(-1)^{a\oplus b}$ for $a,b,x,y \in \set{0,1}$, it is known that $I_0 = 2$, but $I = 2\sqrt{2}$ is possible for a quantum system.

\section{Fresh Randomness from Untrusted Devices} \label{sec:NoSideInfo}

In this section, we recall (some of) the findings of~\cite{nature10}, and also discuss and fix some subtle issue that got neglected there. Throughout this and the upcoming sections, we consider fixed finite sets $\A,\B,\X,\Y$, and a fixed set ${\cal C} = \set{c_{abxy}}$ of  Bell coefficients. The reader may think of CHSH, but our results hold generally.

\subsection{A Single Interaction}\label{sec:SingleRep}

We consider an untrusted device $\D$, prepared by an adversary Eve. As discussed in the introduction, $\D$~consists of two components,\footnote{The results derived here apply also to devices with three or more components, including the three-component devices used by~\cite{CK11}.} which, on respective inputs $x \in \X$ and $y \in \Y$, produce respective outputs $a \in \A$ and $b \in \B$ {\em without communicating}. 
Formally, $\D$'s behavior is given by an unknown conditional
probability distribution $P_{AB|XY}$, which is specified by an unknown
quantum state $\rho_{\ca\cb} \in \dens(\H_{\ca} \otimes \H_\cb)$ 
of unknown dimension, and unknown families of measurements $\{M_x^a\}$ and~$\{N_y^b\}$, acting on the respective systems $\ca$ and $\cb$. We are interested in the guaranteed amount of uncertainty in $A$ and $B$ (conditioned on $X$ and $Y$), under the promise that $P_{AB|XY}$ has some given Bell value, greater than~$I_0$. 
This motivates the following definition.

\begin{definition}
For a given set of Bell coefficients, we define $\h$ to be the function
$$
\h(I) = \inf_{\H_\ca, \H_\cb, \rho_{\ca\cb} \atop \{M_x^a\},\{N_y^b\}} \min_{x \in \X \atop y \in \Y} \; \Hmin(AB\mid X\!=\!x,Y\!=\!y)
$$
where the outer infimum is over all finite dimensional Hilbert spaces
$\H_\ca$ and $\H_\cb$, all states $\rho_{\ca\cb} \in \dens(\H_{\ca}
\otimes \H_\cb)$, and all families of measurements $\{M_x^a\}$ and
$\{N_y^b\}$ such that the resulting conditional probability
distribution \smash{$P_{AB\mid XY}(a,b\mid x,y) = \tr(M_x^a \otimes
  M_y^b \, \rho_{\ca\cb} \, {M_x^a}^\dag \otimes {M_y^b}^\dag)$} has
Bell value at least~$I$.  Also, we define $\clh$ to be the
\emph{convex closure} of $\h$, i.e., the maximal convex function that
does not exceed $\h$.%
\footnote{Formally, $\clh(I) = \max f(I)$ where the maximum is over
  all convex functions $f$ which are upper bounded
  by~$\h$.}
\end{definition}

Pironio \etal~\cite{nature10} show that by means of a hierarchy of
semi-definite programs (SDPs) \cite{NPA07,NPA08}, $\h(I)$~can be
numerically computed up to arbitrary precision (by means of a possibly
expensive computation).  They also show an analytical lower bound of
$1-\log\bigl( 1+\sqrt{2-I^2/4} \bigr)$ for $\h(I)$ in the case of
CHSH, which reaches $1$ for $I = \Imax = 2 \sqrt{2}$ (whereas the
numerical calculation gives $\h(2\sqrt{2}) \approx 1.23$), and
monotonically decreases to $0$ as $I$ goes down to $I_0 = 2$; see
Figure~2 in~\cite{nature10}.  
Since this lower bound is convex, it is also a lower bound on $\clh$;%
\footnote{Actually, the numerical computations for CHSH suggest that $\clh = \h$; we do not know if this holds generally. }
we will need this later on. For now, we can conclude that if an unknown bipartite quantum
system (with fixed measurements $\{M_x^a\}$ and $\{N_y^b\}$) is
promised to have a CHSH value of $I = 2 \sqrt{2}$, then the joint
min-entropy in the measurement outcomes $A$ and $B$ is lower bounded
by approximately $1.23$ bits (respectively $1$ bit, if one wants to
rely on the analytical bound).

\subsection{Sequential Repetitions}\label{sec:SeqRep}

In order to get more uncertainty, and in order to be able to {\em estimate} the Bell value, we consider a sequential repetition of extracting uncertainty from an untrusted device $\D$ as above. 
Informally, rather than interacting with $\D$ once (i.e., inputting
$(x,y) \in \X \times \Y$ and observing $(a,b) \in \A \times \B$),
$\D$~is interacted with $n$ times in sequence, by inputting $(x_1,y_1)
\in \X \times \Y$ and observing $(a_1,b_1) \in \A \times \B$,
inputting $(x_2,y_2) \in \X \times \Y$ and observing $(a_2,b_2) \in \A
\times \B$, etc. This procedure is formalized as follows. 

\paragraph{Modeling. }
We consider an arbitrary but fixed bipartite state $\rho_{\ca \cb} \in \dens(\H_{\ca} \!\otimes\! \H_\cb)$ of an arbitrary finite-dimensional bipartite quantum system $\ca \cb$, and a sequence of $n$ arbitrary but fixed pairs of families of measurements $(\{M_{x_1}^{a_1}\},\{N_{y_1}^{b_1}\}),\ldots,(\{M_{x_n}^{a_n}\},\{N_{y_n}^{b_n}\})$. 

For each pair, \smash{$\{M_{x_j}^{a_j}\}$} is a family of measurements, indexed by $x_j \in \X$, acting on $\ca$, with measurement outcomes $a_j \in \A$, and similar for \smash{$\{N_{y_j}^{b_j}\}$}. 
We allow the two components of the device~$\D$ to communicate {\em
  between} the rounds; this is captured by a sequence $U_2,\ldots,U_n$
of unitary transformations acting on $\H_{\ca} \otimes \H_\cb$, where $U_j$ is
applied to the (collapsed) state before the $j$th interaction. 
For $j\in\{1,\ldots,n\}$, denote with $a^j$ the concatenation of the
first $j$ rounds $a^j=a_1\cdots a_j$ and the same for $b,x$ and
$y$. Let $A^j,B^j,X^j,Y^j$ be the corresponding random variables. To
ease notation, we use bold letters as shortcuts for the concatenation
of all $n$ rounds, e.g. $\ba=a^n,\bA=A^n$, etc.

Formally, the conditional probability distribution $P_{\bA \bB|\bX \bY}$ is defined as 
$$
P_{\bA \bB|\bX \bY}(\ba , \bb\mid\bx,\by) = \prod_{j=1}^n P_{A_j B_j|X_j Y_j \Hist_j}(a_j,b_j\mid x_j,y_j,\hist_{j})
$$
where $\Hist_j = (A^{j-1},B^{j-1},X^{j-1},Y^{j-1})$ and $\hist_j = (a^{j-1},b^{j-1},x^{j-1},y^{j-1})$, and  
$$
P_{A_j B_j|X_j Y_j Hist_j}(a_j,b_j\mid x_j,y_j,hist_j) =
\tr\bigl( (M_{x_j}^{a_j} \otimes N_{y_j}^{b_j}) \,
\rho_{\ca\cb|\Hist_j=\hist_j} \, (M_{x_j}^{a_j} \otimes N_{y_j}^{b_j})^\dag \bigr)
$$
where $\rho_{\ca\cb|\Hist_j=\hist_j}$ is inductively defined for $j =
1,\ldots,n$ as follows. $\rho_{\ca\cb|\Hist_1=\hist_1} = \rho_{\ca
  \cb}$, and, for $1 \leq j < n$,
\begin{align*}
\rho_{\ca\cb|\Hist_{j+1}=\hist_{j+1}} = U_{j+1} \; \frac{(M_{x_{j}}^{a_{j}}
\otimes N_{y_j}^{b_j}) \;
\rho_{\ca\cb|\Hist_j=\hist_j} \; (M_{x_j}^{a_j} \otimes
N_{y_j}^{b_j})^\dag}{P_{A_j B_j|X_j Y_j Hist_j}(a_j,b_j\mid x_j,y_j,hist_j)}  \; U_{j+1}^\dag
\end{align*}
is the state obtained by
applying $U_{j+1}$ to the state to which
$\rho_{\ca\cb|\Hist_{j}=\hist_{j}}$ collapses when $\ca$ and $\cb$ are
measured by $\{M_{x_{j}}^{a_{j}}\}$ and $\{N_{y_{j}}^{b_{j}}\}$,
respectively, and $a_{j}$ and $b_{j}$ are observed.

What is important to realize is that before every round $j$, the
situation is exactly as in the previous Section~\ref{sec:SingleRep},
with a fixed state $\rho_{\ca\cb|\Hist_j=\hist_j}$ and fixed
measurements \smash{$\{M_{x_j}^{a_j}\}$} and
\smash{$\{N_{y_j}^{b_j}\}$} in the device~$\D$, and thus $P_{A_j B_j|X_j Y_j \Hist_j}(\cdot,\cdot|\cdot,\cdot,\hist_{j})$ here behaves as $P_{A B|X Y}$ does in Section~\ref{sec:SingleRep}. 

We would like to point out that there is no need to make $\{M_{x_j}^{a_j}\}$ (and the same for $\{N_{y_j}^{b_j}\}$) dependent on previous in- and outputs, i.e., on $\hist_j$ using the above notation, because we may assume that the measurement $\{M_{x_j}^{a_j}\}$ encodes $x_j$ and $a_j$ into the post-measurement state of $\ca$, and that the subsequent unitary~$U_{j+1}$ copies this (classical) information into the state of $\cb$. The subsequent measurements can then be control measurements, which performs a measurement depending on the history. Similarly, we may assume the $\{M_{x_j}^{a_j}\}$'s to be identical for different $j$'s (and the same for \smash{$\{N_{y_j}^{b_j}\}$}'s), since the quantum system $\ca$ may maintain a counter that is increased by every unitary $U_j$, and $\{M_{x_j}^{a_j}\}$ can then be chosen as a control measurement that is controlled by the counter.%
\footnote{These observations on the independence of the measurements on the history and the round are not crucial for our proofs; they merely simplify the notation. }

Given the conditional probability distribution $P_{\bA \bB|\bX \bY}$
as specified above, which describes the input-output behavior of the
$n$ sequential interactions with the device~$\D$, once a distribution~$P_{\bX \bY}$ is decided upon, which specifies how the inputs $x_j$ and $y_j$ are chosen in each round, 
the joint probability distribution $P_{\bA \bB \bX \bY}$ is determined as $P_{\bA \bB \bX \bY} = P_{\bX \bY} P_{\bA \bB|\bX \bY}$.

\paragraph{Estimating the Bell value. }

Once the device $\D$ is given, i.e.\@ the state $\rho_{\ca \cb}$, the measurements $(\{M_{x_1}^{a_1}\},\{N_{y_1}^{b_1}\})$, $\ldots$, $(\{M_{x_n}^{a_n}\},\{N_{y_n}^{b_n}\})$ and the unitaries $U_2,\ldots,U_n$ are fixed, $P_{A_1 B_1|X_1 Y_1}$ and thus the Bell value of the first round of interaction, $I_1 = I(P_{A_1 B_1|X_1 Y_1})$, is determined. For the other rounds, this is slightly more subtle. The reason is that the state $\rho_{\ca\cb|\Hist_{2}=\hist_{2}}$ before the second round, and thus the probability distribution $P_{A_2 B_2|X_2 Y_2, \Hist_2=\hist_2}$, depends on what happened in the first round, i.e., depends on $\hist_2 = (a_1,b_1,x_1,y_1)$. Thus, the Bell value of the second round, $I_2 = I(P_{A_2 B_2|X_2 Y_2, \Hist_2=\hist_2})$, is a function of $\hist_2$. Similarly, the  Bell value of the $j$-th round, $I_j = I(P_{A_j B_j|X_j Y_j, \Hist_j=\hist_j})$, is a function of $\hist_j$. 
We let 
\begin{equation}\label{eqn:bell}
\bar{I} = \frac{1}{n} \sum_{j=1}^n I_j
\end{equation}
be the average Bell value, averaged over the $n$ rounds, and we write $\bar{I} = \bar{I}(\ba,\bb,\bx,\by)$ to make its dependency on the $\ba$, $\bb$ etc.\@ explicit.%
\footnote{Actually, it only depends on $(a^{n-1},b^{n-1},x^{n-1},y^{n-1})$.}

Pironio \etal show in~\cite{nature10} that the average Bell value $\bar{I}$ can be estimated by analyzing the data collected over the $n$ rounds. Specifically, defining
\begin{equation}\label{eqn:bellEst}
\hat{I} = \hat{I}(\ba,\bb,\bx,\by) = \frac1n \sum_{j=1}^n\sum_{abxy} c_{abxy} \frac{\chi(a_j=a,b_j=b,x_j=x,y_j=y)}{P_{XY}(x,y)}
\end{equation}
where $\chi(e)$ is the indicator of the event $e$ (that is, $\chi(e)=1$ if the event $e$ occurs and 0 otherwise), the following holds.
 
\begin{proposition}[\cite{nature10}]\label{prop:n-estimate}
For $\bar{I}$ and $\hat{I}$ as above, for arbitrary but iid $(\bX,\bY)$, meaning that $P_{\bX \bY} = \prod_j P_{X_j Y_j}$ with $P_{X_j Y_j} = P_{XY}$ for all $j$, 
and for any $\eps > 0$: 
\[
P\Bigl[ \bar{I}(\bA,\bB,\bX,\bY)  \leq \hat{I}(\bA,\bB,\bX,\bY) -\eps \Bigr] \leq \exp\Biggl(- \frac{\eps^2 n}{2(\frac{c_{\max}}{p_{\min}}+\Imax)} \Biggr) \, ,
\]
where $\Imax$ is the maximal value of $I$ achievable by means
of a quantum system, and $p_{\min} = \min_{x,y} P_{XY}(x,y)$ and $c_{\max} = \max\set{c_{abxy}}$.\end{proposition}

Thus, except with small probability, the estimated value $\hat{I}$ for the average Bell value is not much smaller than the real average Bell value $\bar{I}$. 
For a fixed choice of Bell coefficients ${\cal C} = \set{c_{abxy}}$, which uniquely determines $\Imax$, we write $c(p_{\min}) = \frac{\ln 2}{2}(\frac{c_\text{max}}{p_{\min}}+\Imax)^{-1}$, so that the probability in Proposition~\ref{prop:n-estimate} can be written as \smash{$2^{-c(p_{\min}) \eps^2 n}$}. 

We stress that for Proposition~\ref{prop:n-estimate} to hold, it is crucial that $\bX$ and $\bY$ are chosen {\em independently} of the internal state of $\D$; this is implicit in the statement of Proposition~\ref{prop:n-estimate} by having modeled the internal state~$\rho_{\ca \cb}$ of~$\D$ to be fixed and independent of $\bX$ and~$\bY$: $\rho_{\bX \bY \ca \cb} = \rho_{\bX \bY}  \otimes \rho_{\ca \cb} $. Obviously, if $\D$ knows $\bX$ and $\bY$ in advance, then it can easily pretend to have a large Bell value while, for instance, being classical.

\paragraph{Bounding the min-entropy. }

It remains to argue that if $\bar{I}$ is non trivial, i.e.\@ sufficiently greater than~$I_0$, which can be learned by observing $\hat{I}$ (except with small probability), then the pair $(\bA,\bB)$ contains a linear (in $n$) amount of min-entropy. 
To this end, Pironio~\etal show (see equation (A.5) in~\cite{nature10}) that 
\begin{equation}\label{eq:bound}
P_{\bA\bB\mid \bX \bY}(\ba,\bb\mid \bx,\by) \leq 2^{- n\cdot \clh(\bar{I}(\ba,\bb,\bx,\by))}
\end{equation}
for all $\ba,\bb,\bx,\by$. In the derivation, they use the fact 
that $\clh$ is convex. 
From~\eqref{eq:bound}, they conclude (see equation (A.9) in~\cite{nature10}) that $\hmin(\bA \bB\mid \bX\!=\!\bx,\bY\!=\!\by) \geq n\cdot \clh(\bar{I})$ and thus $\geq n\cdot \clh(\hat{I}-\eps)$ except with small probability. 
However, this conclusion does not seem correct. What follows from
(\ref{eq:bound}) is that 
\begin{equation}\label{eq:derivedbound}
\hmin(\bA \bB\mid \bX\!=\!\bx,\bY\!=\!\by) \geq n\cdot
\clh(\bar{I}(a_\circ^n,b_\circ^n,\bx,\by))
\end{equation}
for the values $a_\circ^n$ and $b_\circ^n$ that {\em minimize} the
right-hand side of~\eqref{eq:derivedbound}; but then, the right-hand
side of~\eqref{eq:derivedbound} is likely to be smaller than $n\cdot \clh(\bar{I}(\ba,\bb,\bx,\by))$ or $n\cdot \clh(\hat{I}(\ba,\bb,\bx,\by))$ for the  values $\ba$ and $\bb$ actually {\em observed}.%
\footnote{When approached with this issue, the authors of~\cite{nature10} confirmed
that their formulation is improper, and they mentioned that they have
been aware of it and know how to solve it. In particular, their
independent work~\cite{PM11} fixes this issue in a similar manner as we do
here.}

For the remainder of this section, we propose and discuss a possible
way to get a meaningful and useful statement on the min-entropy of
$(\bA,\bB)$ in terms of $\clh(\bar{I})$, and thus of $\clh(\hat{I})$
except with small probability.  We partition the interval
$[I_0,I_\text{max}] \subset \R$, ranging from the trivial---meaning
classical---Bell value $I_0$ to the maximal value $I_\text{max}$, into
$m$ disjoint blocks: $[I_0,I_\text{max}] = \Omega_0 \cup \ldots \cup
\Omega_{m-1}$, where $\Omega_\ell$ is of the form $\Omega_\ell =
[J_\ell,J_{\ell+1})$, with the exception that $\Omega_{m-1} =
[J_{m-1},I_\text{max}]$, for some boundary points $I_0=J_0 \leq J_1 \leq \cdots \leq J_{m-1}
\leq I_\text{max}$. The value of $m \in \N$ and
the (possibly different) sizes of the $\Omega_\ell$'s are arbitrary
but fixed. 

For any parameter $\eps > 0$, given the random variables $\bA,\bB,\bX,\bY$, describing
the $n$ interactions with the device $\D$, we can now define the random
variable $L_\eps$ to be the unique random variable that satisfies $\hat{I}(\bA,\bB,\bX,\bY) -\eps \in \Omega_{L_\eps}$ (with natural adjustments outside of the range $[I_0,I_\text{max}]$).%
\footnote{The definition of $L_\eps$ simply captures that if $\hat{I}$ is too close to the lower end of an interval, then we take the next lower interval to be on the safe side. } 

\begin{theorem}\label{thm:noSideInfo}
Let $(\bX,\bY)$ be iid. Then, for any $\eps,\delta > 0$ there
exists a ``good'' event $\GOOD$ with $P[\GOOD] \geq 1 - m \cdot
2^{-\delta n} - 3 \cdot 2^{-c(p_{\min})\eps^2 n}$, and such that 
$$
\Guess(\bA \bB\mid\bX\!=\!\bx, \bY\!=\!\by, L_\eps\!=\!\ell, \GOOD) \leq 2^{-n \cdot \clh(J_\ell) + \delta n + 1}
$$
and thus
$$
\Hmin(\bA \bB\mid\bX\!=\!\bx, \bY\!=\!\by, L_\eps\!=\!\ell, \GOOD) \geq n \cdot \clh(J_\ell) - \delta n - 1 
$$
for all $\bx \in \X^n$, $\by \in \Y^n$ and $\ell \in \set{0,\ldots,m-1}$ with $P_{\bX \bY L_\eps|\GOOD}(\bx,\by,\ell) > 0$.
\end{theorem}
We would like to point out that for the bound on $P[\GOOD]$ to hold,
it is crucial that $\rho_{\ca \cb}$ is independent of $(\bX,\bY)$ (and
the $(X_i,Y_i)$'s are iid): 
clearly, the device can fool you if it
knows the inputs it will get in advance. However, for the event
$\GOOD$ as defined in the proof below, the bound on the guessing
probability holds {\em irrespectively} of the distribution of $\bX$
and $\bY$.  Indeed, the value of $\Guess(\bA \bB\mid\bX\!=\!\bx,
\bY\!=\!\by, L_\eps\!=\!\ell, \GOOD)$ is determined by the
conditional probability distribution $P_{\bA \bB|\bX
  \bY}(\cdot,\cdot|\bx,\by)$ alone (which is determined by $\rho_{\ca \cb}$, the family of measurements and the unitaries); this holds because $L_\eps$
as well as $\GOOD$ (this, we will see below) are uniquely determined
by $\bA,\bB,\bX$ and $\bY$.

\begin{proof}
  Let $\BADGUESS$ be the bad event $\bar{I}(\bA,\bB,\bX,\bY) \leq
  \hat{I}(\bA,\bB,\bX,\bY) -\eps$ that the estimated Bell value $\hat{I}$ is
  significantly larger than the average Bell value $\bar{I}$, and let
  $\GOODGUESS$ be its complement (which we understand as a good event); by
  Proposition~\ref{prop:n-estimate}, we know that $P[\BADGUESS] \leq
  2^{-c(p_{\min}) \eps^2 n}$. We define $\BAD_1$ to be the set of all
  ``bad inputs'' $(\bx,\by)$ with the property that
$$
P[\BADGUESS|\bX=\bx,\bY=\by] \geq \frac12 \, ;
$$ 
it is straightforward to show that $P[(\bX,\bY) \in \BAD_1] \leq 2\cdot2^{-c(p_{\min}) \eps^2 n}$. 
Finally, we define $\BAD_2$ to be the set of all $(\bx,\by,\ell)$ with the property that 
$$
P_{L_{\eps}|\bX \bY \GOODGUESS}(\ell|\bx,\by) \leq 2^{-\delta n} \, .
$$
It follows from the definition of $\BAD_2$ that
$P[(\bX,\bY,L_{\eps}) \in \BAD_2 | \GOODGUESS] \leq m \cdot
2^{-\delta n}$. We slightly abuse
notation and identify the {\em set} $\BAD_1$ with the bad {\em event}
$(\bX,\bY) \in \BAD_1$ and we write $\GOOD_1$ for its complementary
good event, and correspondingly for $\BAD_2$ and $\GOOD_2$. We now
define the good event $\GOOD$ as $\GOOD := \GOODGUESS \wedge \GOOD_1
\wedge \GOOD_2$. Using union bound over the bad events, it is not too hard to show that 
$P[\GOOD] \geq 1 - m \cdot 2^{-\delta n} - 3\cdot2^{-c(p_{\min})\eps^2 n}$.

It remains to argue the bound on the min-entropy. Let
$\ba,\bb,\bx,\by$ be such that $\bar{I}(\ba,\bb,\bx,\by)  >
\hat{I}(\ba,\bb,\bx,\by) -\eps$, i.e., they have positive probability
conditioned on the good event $\GOODGUESS$. Furthermore, let
$\ell$ be the unique value with $\hat{I}(\ba,\bb,\bx,\by) -\eps \in
\Omega_{\ell}$. 
If $(\bx,\by) \not\in \BAD_1$, then $P[\GOODGUESS|\bX=\bx,\bY=\by] \geq
\frac12$ and hence, conditioning on the event $\GOODGUESS$ can
increase the probabilities by at most a factor of 2. For those
$(\bx,\by) \not\in \BAD_1$, it then follows from (\ref{eq:bound}) that
$$
P_{\bA\bB\mid \bX \bY, \GOODGUESS}(\ba,\bb\mid \bx,\by) \leq 2 \cdot 2^{- n\cdot \clh(\bar{I}(\ba,\bb,\bx,\by))} \leq 2 \cdot 2^{- n\cdot \clh(\hat{I}(\ba,\bb,\bx,\by)-\eps)} \leq 2 \cdot 2^{- n\cdot \clh(J_\ell)} \, .
$$
If additionally we have $(\bx,\by,\ell) \not\in \BAD_2$, then
$$
P_{\bA\bB\mid \bX \bY L_\eps, \GOODGUESS}(\ba,\bb\mid \bx,\by,\ell) \leq 
\frac{P_{\bA\bB\mid \bX \bY, \GOODGUESS}(\ba,\bb\mid \bx,\by)}{P_{L_\eps \mid \bX \bY  \GOODGUESS}(\ell\mid \bx,\by)}
\leq 2 \cdot 2^{- n\cdot \clh(J_\ell)} \cdot 2^{\delta n} \, .
$$
Note that additionally conditioning on $\GOOD_1$ and $\GOOD_2$ does not change the above conditional probability distribution if $(\bx,\by) \not\in \BAD_1$ and $(\bx,\by,\ell) \not\in \BAD_2$. Thus, the same bound also applies to $P_{\bA\bB\mid \bX \bY L_\eps, \GOOD}(\ba,\bb\mid \bx,\by,\ell)$, for all $\ba,\bb,\bx,\by$ and $\ell$ with $P_{\bA \bB \bX \bY L_\eps|\GOOD}(\ba, \bb, \bx,\by,\ell) > 0$. 
By definition of the guessing probability and the min-entropy, this proves the claim. 
\end{proof}

\paragraph{A specific example. } 
Consider CHSH, so that the Bell value of a given device is expected to be in the range from $I_0 = 2$ to $\Imax = 2\sqrt{2} \approx 2.828$. Let us divide this range into $J_0 = I_0 < J_1 = 2.2 < J_2 = 2.4 < J_3 = 2.6 < \Imax$, and let us take a $q$-biased input distribution $P_{\bX \bY} = \prod_j P_{X_j Y_j}$ with $P_{X_j Y_j}(0,0) = 1-3q$ and $P_{X_j Y_j}(x,y) = q$ for all $(x,y)\in \set{0,1}^2 \setminus \set{(0,0)}$, where $0 < q \leq 1/4$ is some parameter. Finally, let us fix some small parameters $\eps,\delta > 0$; for concreteness, say that $\eps = 0.05$ and $\delta = 0.01$.

Consider now $n$ sequential interactions with an untrusted device
$\D$, where in each round $x_j$ and $y_j$ are chosen (according to
$P_{X_j Y_j}$) and input into $\D$, and $a_j$ and $b_j$ are obtained
as output from $\D$. Let us say that from the collected data, we get
$\hat{I}(\ba,\bb,\bx,\by) = 2.7 \in \Omega_3$ as estimation for the
average Bell value. By Theorem~\ref{thm:noSideInfo}, he have that
given $\bx$ and $\by$ and $L_\eps = 3$, the min-entropy of $\ba$ and
$\bb$ is at least $n \cdot (\clh(2.6) - \delta) - 1 \approx n \cdot
(0.36 - \delta) > n/3$ bits, except with probability $4 \cdot
2^{-\delta n} + 3\cdot2^{-c(q)\eps^2 n}$.%
\footnote{This probability is {\em on average} over the execution;
  given a specific outcome for $\hat{I}(\ba,\bb,\bx,\by)$, like $2.7$
  here, the probability may be different. } Thus, when applying a
suitable randomness extractor to $\ba,\bb$, we can extract, say, $n/4$
bits that are exponentially close to uniformly distributed (given
$\bx$ and $\by$ and $L_\eps = 3$).

In order to sample the inputs according to the biased input
distribution $P_{\bX \bY}$, as suggested in~\cite{nature10}, it is
known to be sufficient (in average) to have access to $n \cdot O(q
\log(1/q))$ random bits~\cite{KY76}. Since, $q \log(1/q)$ converges to
$0$ for $q \rightarrow 0$, if $q$ is chosen to be a small enough
constant, then, say, $n/4$ random bits are sufficient. Thus, by
starting off with $n/4$ random bits,%
\footnote{We are ignoring here the randomness needed for the
  extractor.}  we obtained another $n/4$ almost-random bits and thus
hold now $n/2$ random bits. Thus, we have expanded the randomness by a
factor $2$. Choosing $q = O(1/\sqrt{n})$, one obtains an expansion
factor $O(\sqrt{n}/\log n)$ while still being negligibly close to
perfect randomness (since $c(1/\sqrt{n}) = \Omega(1/\sqrt{n})$).

Having generated fresh randomness from an untrusted device $\D$, one
is now tempted to use the newly obtained randomness to generate even
more fresh randomness from the device~$\D$, and so on. This does not
work. The reason is that the generated randomness is not random {\em
  to the device}~$\D$, or, more formally, not independent of the
internal state of~$\D$; indeed, $\D$ has already observed $\bx$ and
$\by$ and it has itself produced $\ba$ and $\bb$. We argue below,
however, that we can use the fresh randomness to generate even more
randomness from {\em another} device, as long as the devices are not
entangled with each other nor with the adversary.

\paragraph{Classical Side Information}
The case where the adversarial producer Eve of the devices holds classical
side information about the device $\D$, can be reduced to the case without
side information by conditioning on particular values of the side
information.

\section{Composability}\label{sec:composability}
Consider two (or more) untrusted devices $\D$ and~$\D'$, prepared by
the adversary Eve. We assume that $\D$~and~$\D'$ cannot communicate
and are not entangled with each other. The case when Eve holds
classical side information about the devices can be treated as
described in the previous section. We can then apply
Theorem~\ref{thm:noSideInfo} to argue that the output $\bA \bB$
produced by $\D$ has high min-entropy (except with small probability)
given the internal state of~$\D'$ (because $\D'$ is independent of
$\D$), assuming that a large enough average Bell value is observed. It
thus follows that by applying an extractor (with suitable parameters
and a freshly chosen seed) to $\bA \bB$, we obtain a bitstring~$K$
that is close to random and independent of the internal state
of~$\D'$. This in particular implies that if we use the randomness~$K$
to sample the input $\bX' \bY'$ to $\D'$ (according to a prescribed
distribution), then $\bX' \bY'$ is close to {\em independent} of the
internal state of~$\D'$. As the dependency between the internal
(quantum) state of $\D$ and the in-/outputs of $\D'$ is purely
classical, we can condition on this classical information and
apply Theorem~\ref{thm:noSideInfo} to argue that
the output~$\bA' \bB'$ produced by~$\D'$ has high min-entropy given
the current internal state of~$\D$.  Therefore, we are in the
same situation as above, and so can use the randomness extracted from
$\bA' \bB'$ to sample again inputs for~$\D$, and we can keep on going
like this as long as a large enough Bell value is observed. We stress
that for the above line of reasoning only works because we assumed the
devices $\D,\D'$ to be unentangled to start with.
In order to see quantitatively how this procedure can lead to a
superpolynomial randomness expansion, we refer to~\cite[Section~5]{FGS11v2}.

\section{Conclusion and Open Problems}
An interesting extension to our result is to generalize
Theorem~\ref{thm:noSideInfo} to the setting of quantum side information.
This would allow a composition theorem for the more general case
in which the devices can be entangled with each other and with Eve.
Numerical calculations seem to suggest that the bound on the min-entropy does carry over to the quantum setting. 
%
Unfortunately, we are unable at the moment to give a rigorous proof of this claim and leave it as main open question.

\section*{Acknowledgment}
RG is grateful to CWI,  Amsterdam for hosting him while part of this
work was done. CS is supported by a NWO VENI grant.


\bibliographystyle{alpha}
\bibliography{bib-bell}


\end{document}